\documentclass[10pt]{article}
\usepackage{inputenc}
\usepackage{amsmath}
\usepackage{amssymb, amsfonts}
\usepackage{graphicx,epstopdf,bm}
\usepackage[colorlinks,bookmarks,urlcolor=blue,citecolor=blue,linkcolor=blue]{hyperref}
\usepackage[T1]{fontenc}

\newcommand{\pd}[2]{\frac{\partial #1}{\partial #2}}
\newcommand{\pdd}[2]{\frac{\partial^{2} #1}{\partial #2 ^{2}}}

\newcommand{\abs}[1]{\left\vert #1 \right\vert}
\newcommand{\ave}[1]{\left \langle #1 \right \rangle}

\newcommand{\mum}{\mu \mathrm{m}}
\newcommand{\nm}{\mathrm{nm}}
\newcommand{\vr}{\mathbf{r}}
\newcommand{\vz}{\mathbf{z}}

\newcommand{\norm}[1]{\Vert #1 \Vert}
\newcommand{\dl}{L_0}
\newcommand{\ndl}{l_0}
\newcommand{\dT}{\mathcal{T}}
\newcommand{\ndT}{\bar{T}}
\newcommand{\AN}{A_{N}}
\newcommand{\Cn}[1]{C_{#1}}
\newcommand{\CN}{\Cn{N}}

\begin{document}
\title{First-passage time to clear the way for receptor-ligand binding in a crowded environment}
\author{Jay Newby\thanks{Department of Mathematics, University of North Carolina, Chapel Hill, 329 Phillips Hall, Chapel Hill, NC 27599} \and Jun Allard\thanks{Department of Mathematics, University of California, Irvine, 340 Rowland Hall, Irvine, CA 92697}}

\maketitle

\begin{abstract}
Certain biological reactions, such as receptor-ligand binding at cell-cell interfaces and macromolecules binding to biopolymers, require many smaller molecules crowding a reaction site to be cleared. Examples include the T cell interface, a key player in immunological information processing.
Diffusion sets a limit for such cavitation to occur spontaneously, thereby defining a timescale below which active mechanisms must take over.
We consider $N$ independent diffusing particles in a closed domain, containing a sub-region with $N_{0}$ particles, on average.
We investigate the time until the sub-region is empty, allowing a subsequent reaction to proceed.
The first passage time is computed using an efficient exact simulation algorithm and an asymptotic approximation in the limit that cavitation is rare.
In this limit, we find that the mean first passage time is sub-exponential, $T \propto e^{N_{0}}/N_{0}^2$. 
For the case of T cell receptors, we find that stochastic cavitation is exceedingly slow, $10^9$ seconds at physiological densities, however can be accelerated to occur within 5 second with only a four-fold dilution. 
\end{abstract}


Diffusion drives many biological processes, both positively, by delivering cargo to a target, and negatively, by removal of cargo from a region of interest (ROI). 
While the temporal dynamics of diffusional delivery have been extensively studied \cite{Bressloff:2012vm, Berg:1993ug, Klenin:2006da, Coombs:2009}, diffusion-driven removal has been less characterized experimentally or theoretically \cite{BenNaim:2010}. 
Removal is of particular interest in the crowded environment of cells, where large biomolecules and cellular structures require the displacement of smaller molecules, a phenomenon we term stochastic cavitation. 

\begin{figure}[bp]
  \centering
  \includegraphics[width=8cm]{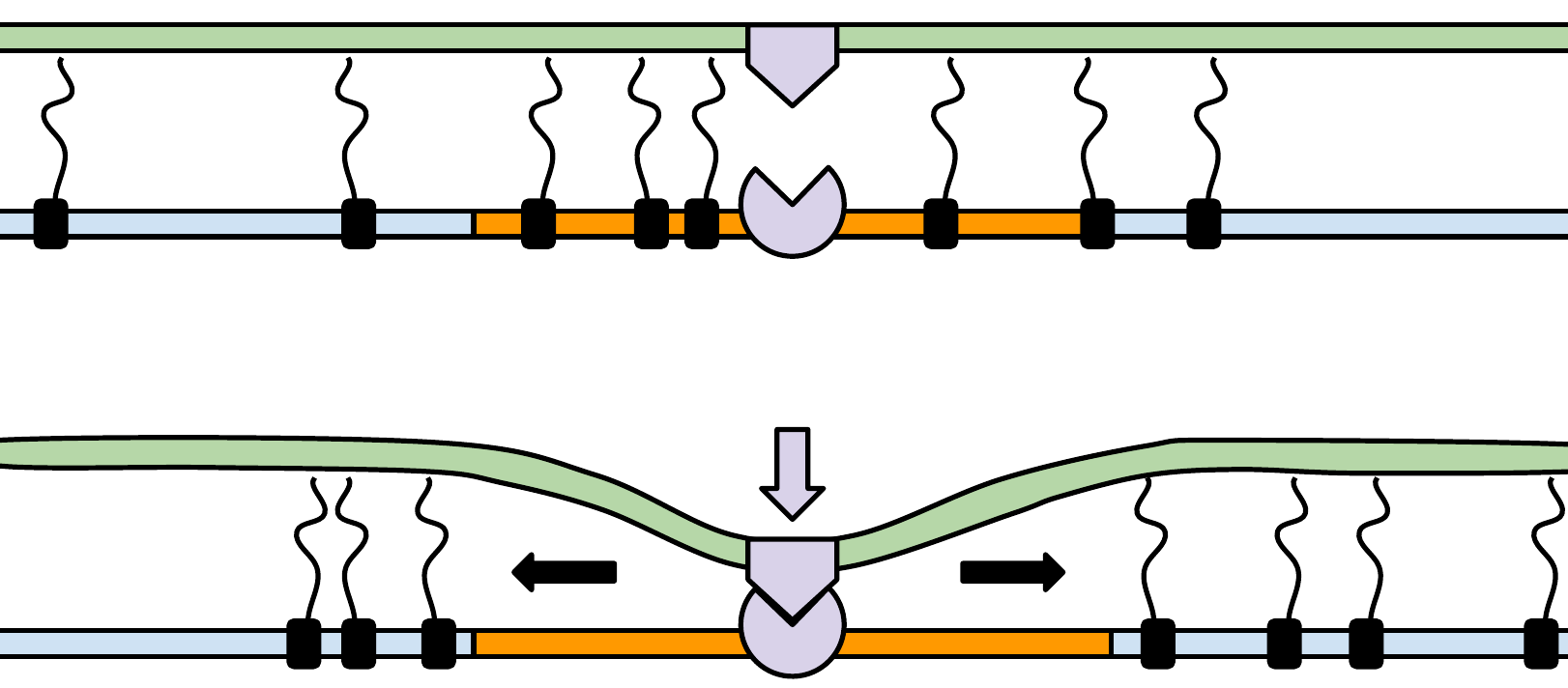}
  \caption{Cell-cell interface formation between a T-cell and an antigen-presenting cell. CD45 molecules (black) block the receptor-ligand (purple) bond from forming while they inhabit the ROI (orange). }
  \label{fig:diag}
\end{figure}
A specific example arises in the study of cell-cell interfaces including the T-cell/antigen-presenting-cell interface \cite{Kaizuka:2007ik, Allard:2012gy, Rozycki:2010gy, Chattopadhyay:2007dd} (see Fig.~\ref{fig:diag}).
A fundamental question for all cell-cell interfaces is how receptors and ligands come into contact, despite being separated by large molecules, the extracellular fluid, and other structures in the glycocalyx.
On either cell surface, large molecules such as CD45 and LFA-1 undergo 2D diffusion in the cell membrane with a diffusion coefficient of $D \sim 0.1 \mum^2/s$ \cite{Rajani:2011jpa, Cairo:2010goa}.
These large molecules impair interactions between smaller pairs of molecules, such as the T cell receptor and its ligand---a key step in immunological information processing and decision-making. 
It has been estimated that a region of radius $\sim100\nm$, devoid of large molecules, is necessary for spontaneous T cell receptor interaction \cite{Allard:2012gy}, which is occupied by on average $\sim30$ particles at equilibrium. A natural question is whether this empty region can form spontaneously in a biologically relevant time.  Understanding contact formation will address cell-cell interactions in the crowded, heterogeneous environment inside organisms and what machinery is necessary for general cell-cell interactions \cite{Goodridge:2011,Shergill:2012}.

Examples of diffusional cavitation in biology also arise in other dimensionalities. In 1D, microtubules (inflexible polymers of the protein tubulin) are decorated by hundreds of microtubule-associated proteins \cite{Alberts:2014dy, Rouzier:2005kg}.
These proteins exhibit significant crowding \cite{Conway:2012cm} and lateral diffusion along the microtubule lattice \cite{Dixit:2008ip, Helenius:2006ik}.
Large microtubule-binding molecules may therefore have to wait for a region to be clear before binding.
What is the mean time for such clearance, and is it the rate-limiting step in microtubule binding? A similar situation occurs for DNA and the myriad of DNA-binding molecules, some of which undergo lateral diffusion across base pairs \cite{Hammar:2012ex, McKinney:2004jr}.
A significant waiting-time for large DNA-binding molecules has potential implications for the study of the chemical modification of DNA and RNA, all of which require an enzyme to attach to the polymer.

For some of the above scenarios, it has been hypothesized that clearance of the target region requires an active process \cite{Allard:2012gy, Hoerndli:2013jb}.
To address the feasibility of passive diffusion-driven cavitation, a theoretical assessment of the timescales involved is needed.
In other words, can diffusion-driven cavitation reliably occur on biologically relevant timescales?
To address this question, we consider $N$ independent particles undergoing simple diffusion in either the 1D domain $(-L, L)$ or the 2D domain $(-L, L)^{2}$.
In each case, we study the first-passage time until a smaller region, a disk of radius $\dl$, is empty.

The cavitation event can be rare (i.e., the first passage time can be very large compared to the diffusion timescale $L^{2}/D$) under certain circumstances.
To understand this, consider the 1D domain with $N$ particles.
At equilibrium, each particle has a uniformly distributed postion within the domain.
In the limit $L\to\infty$ and $N\to\infty$ with the average particle density $\varphi = N/(2L)$ fixed, the equilibrium probability of finding a region of radius $\dl$ containing no particles is small, $P = e^{- N_{0}}$, where $N_{0} = 2L\varphi$.
Therefore, when $N_{0} \gg 1$ we expect cavitation to be a rare event.
Although we might expect the 1D mean first-passage time (MFPT) to scale as\footnote{This problem is equivalent to the diffusion of a particle in $ND$ dimensions (the product of number of particles and dimensionality of space). Since $ND\gg2$, this Brownian motion is not recurrent, so we might naively expect the system to be well-mixed in $ND$-dimensional phase space, and the rate of first passage would be the attempt rate times the probability of being in the target state.} $T\propto P^{-1} \sim e^{N_{0}}$, we instead we find an asymptotic scaling of $T \propto e^{N_{0}}/N_{0}^2$. 

In this Letter, we develop a simulation algorithm to efficiently generate exact realizations of the first passage time, based on Green's function reaction dynamics \cite{oppelstrup09a}.
For situations where cavitation is a rare event and computation becomes unfeasible, i.e., when $N$ is very large or the ROI occupies most of the explorable area, we develop an asymptotic approximation of the mean first passage time.

Consider $N$ independent random walkers $Y_{n}(t)$, with $n = 1, \cdots, N$, that are confined to the interval $-L < y < L$.
The ROI is the inner domain centered at the origin with radius $\dl < L$.
The event we wish to characterize is the first time at which the ROI is empty (i.e.,  $\min_{n}\{ Y_{n}(t) \} = \dl$).
We first nondimensionalize the problem using the space scale $L$ and the time scale $L^{2}/D$, where $D$ is the diffusion coefficient.
We define the nondimensional distances $R_{n} = \abs{Y_{n}}/L$, $\epsilon = 1 - \dl/L$, and $\ndl = \dl/L$.
Then, a given particle is inside the ROI if $0< R_{n} < \ndl$.

By formulating a simulation algorithm, we can generate exact samples of the first passage time.
We take advantage of explicit formulas for the probability distributions that govern single particle Brownian motion in a closed domain.
Note that even though we focus on the 1D and 2D cavitation problem in this letter, exact distributions are also known for 3D Brownian motion \cite{carslaw59a}.
The algorithm proceeds as follows.
Given a set of random starting positions $\{R_{n}(t_{0})\}_{1\leq n \leq N}$, select a particle that is inside the ROI and closest to the origin.
That is, select $R_{m} = \min \{R_{n}\} < \ndl$.
The first step is to compute the {\em first} time $\tau$ at which the selected particle leaves the ROI (i.e., $R_{m}(t_{0} + \tau) = \ndl$).
Once $\tau$ has been computed, set $t' = t_{0} + \tau$.
The cavitation event cannot have occurred before time $t'$ because we are certain that $R_{m}(t) < \ndl $ for all $t_{0}< t < t'$.
Therefore, the position of the other particles between time $t_{0}$ and time $t'$ is irrelevant, we need only generate the random position for each of the remaining particles at time $t'$.
Once all positions have been updated, select a new $R_{m} = \min \{R_{n}(t')\}$.
We know that the cavitation event has occurred if $R_{m} \geq  \ndl$.
If $R_{m}< \ndl$, then set $t_{0} = t'$ and repeat the above procedure.

At each step, the jump times $\tau$ can be sampled from the exact distribution $f(\tau | r_{0})$, obtained from the fundamental solution to the diffusion equation with a reflecting boundary at $r=0$ and an absorbing boundary at $r=\ndl$.
The random positions can be sampled from the distribution $p(r | r_{0}, \tau)$, satisfying the diffusion equation with reflecting boundaries at $r = 0$ and $r = 1$.
An efficient way of sampling from $p$ is to use a rejection method, similar to the one described in \cite{oppelstrup09a}.
For the jump time sampled from $f$, we found that the rejection method could not easily be adapted to our situation.
Instead, we sample the jump time using a root finding algorithm.
Additional details are provided in Supplementary Material. 
The simulation algorithm is maximally fast in the sense that only the (average) slowest particle determines the next event time, allowing us to efficiently access densities around $\phi L_0\sim13$. 
We find this is sufficiently high to validate our asymptotic approximations. 

To obtain a complete picture of cavitation in the rare event limits, we develop an asymptotic approximation for the MFPT, $\ndT$.
The approximation is derived for 1D cavitation, and based on simulations, we observe that in the limit $L\to\infty$ with a fixed particle density, the approximation is also surprisingly accurate for 2D cavitation.
We first state the main results (Eqs.~\ref{eq:3}-\ref{eq:7}) and then summarize their derivation.

\begin{figure}[htbp]
  \centering
  \includegraphics[width=8cm]{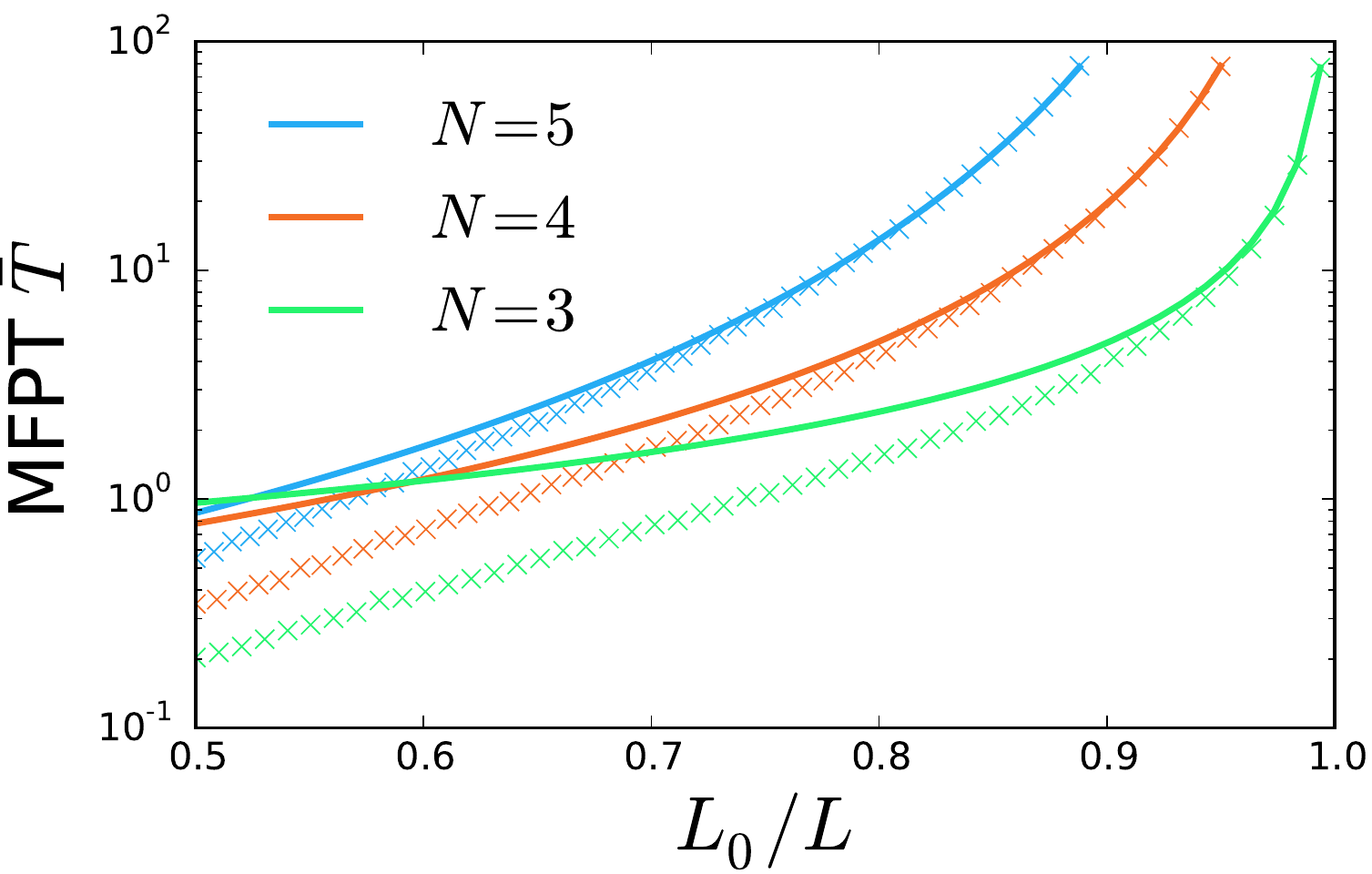}
  \caption{The MFPT (in nondimensional units) as a function of $\dl/L = 1-\epsilon$. The small $\epsilon$ approximation (solid lines) is compared to simulations (symbols), using $\Cn{3} = 1.3214$, $\Cn{4} = 1.44$, and $\Cn{5} = 1.55$.}
  \label{fig:smallEpsilon}
\end{figure}

For fixed $N$, the first term in the asymptotic approximation for $0 < \epsilon \ll 1$ of the MFPT, averaged over a uniformly distributed initial position for each of the $N$ particles, is given by
\begin{equation}
  \label{eq:3}
  \ndT \sim
 \dfrac{2^{N}\AN}{(\CN\epsilon)^{N-2}} + O(1), \quad N \geq 3
\end{equation}
where
\begin{equation}
  \label{eq:2}
  \AN = \frac{\Gamma(\frac{N}{2})}{2\pi^{\frac{N}{2}}(N-2)}.
\end{equation}
The constant $\CN$ is the {\em Newtonian capacitance} of a hypercube in $\mathbb{R}^{N}$; as explained below, it determines the far field behavior of certain solutions to Laplace's equation \cite{spitzer64a}.
An explicit formula for the Newtonian capacitance of a cube for $N>2$ is unknown.
However, a good approximation for $N=3$ is $\Cn{3} \approx 1.3214$ \cite{hwang10a}.
The $\epsilon \ll 1$ approximation (solid line) is compared to simulations (symbols) in Fig.~\ref{fig:smallEpsilon}.
For $N=3$, we find good agreement between simulation and the independently derived estimate for $\Cn{3}$ from \cite{hwang10a}.
From physical arguments detailed at the end of this letter, we have determined an expansion of the Newtonian capacitance for large $N$ given by
\begin{equation}
  \label{eq:75}
    \CN \sim\sqrt{\frac{2N}{\pi e}}\left(1 + \frac{3\log N}{2N} + \frac{\alpha_{2} }{N} + O(N^{-2})\right).
\end{equation}
The unknown constant in the above expansion is independent of all parameters.
Using the exact simulation algorithm, we obtain the numerical estimate, $\alpha_{2} \approx -1.67$.
Our MFPT calculation thus provides an approximation for the capacitance $C_{N}$, which otherwise remains  challenging to compute \cite{hwang10a}.

\begin{figure}[htbp]
  \centering
  \includegraphics[width=8cm]{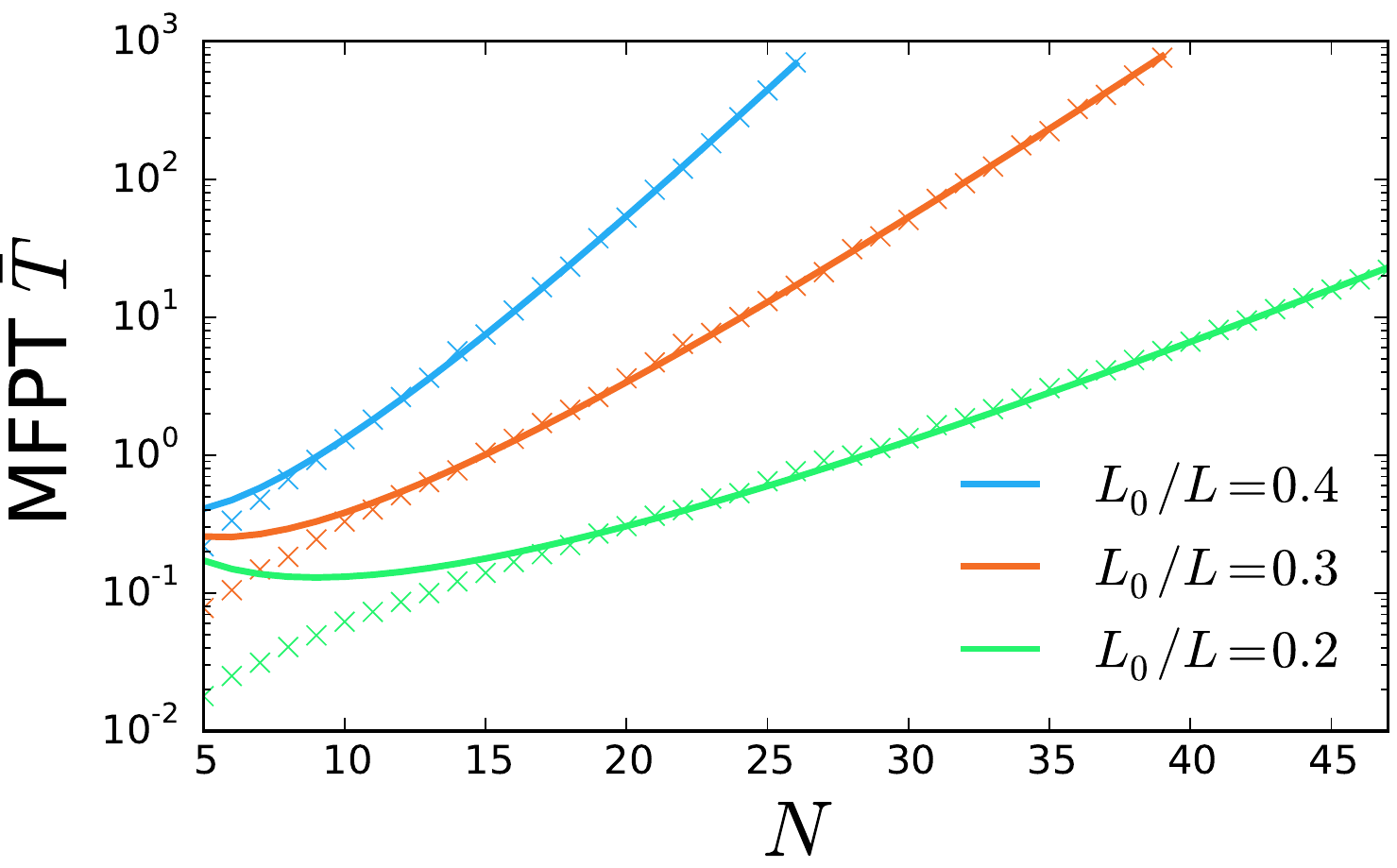}
  \caption{The large $N$ MFPT approximation (solid lines) compared to simulations (symbols).}
  \label{fig:largeN}
\end{figure}

For fixed $0<\epsilon<1$, an asymptotic expansion for $N\gg 1$ is given by
\begin{equation}
  \label{eq:23}
    \ndT \sim \frac{\kappa_{\rm 1D} }{N^{2}\epsilon^{N-2}}, \quad N\gg 1,
\end{equation}
where $\kappa_{\rm 1D} \approx 2.2$ depends only on $\alpha_{2}$ (via Eq.~\ref{eq:4bjun}).
The $N \gg 1$ MFPT approximation is compared to simulations in Fig.~\ref{fig:largeN}.

\begin{figure}[htbp]
  \centering
  \includegraphics[width=8cm]{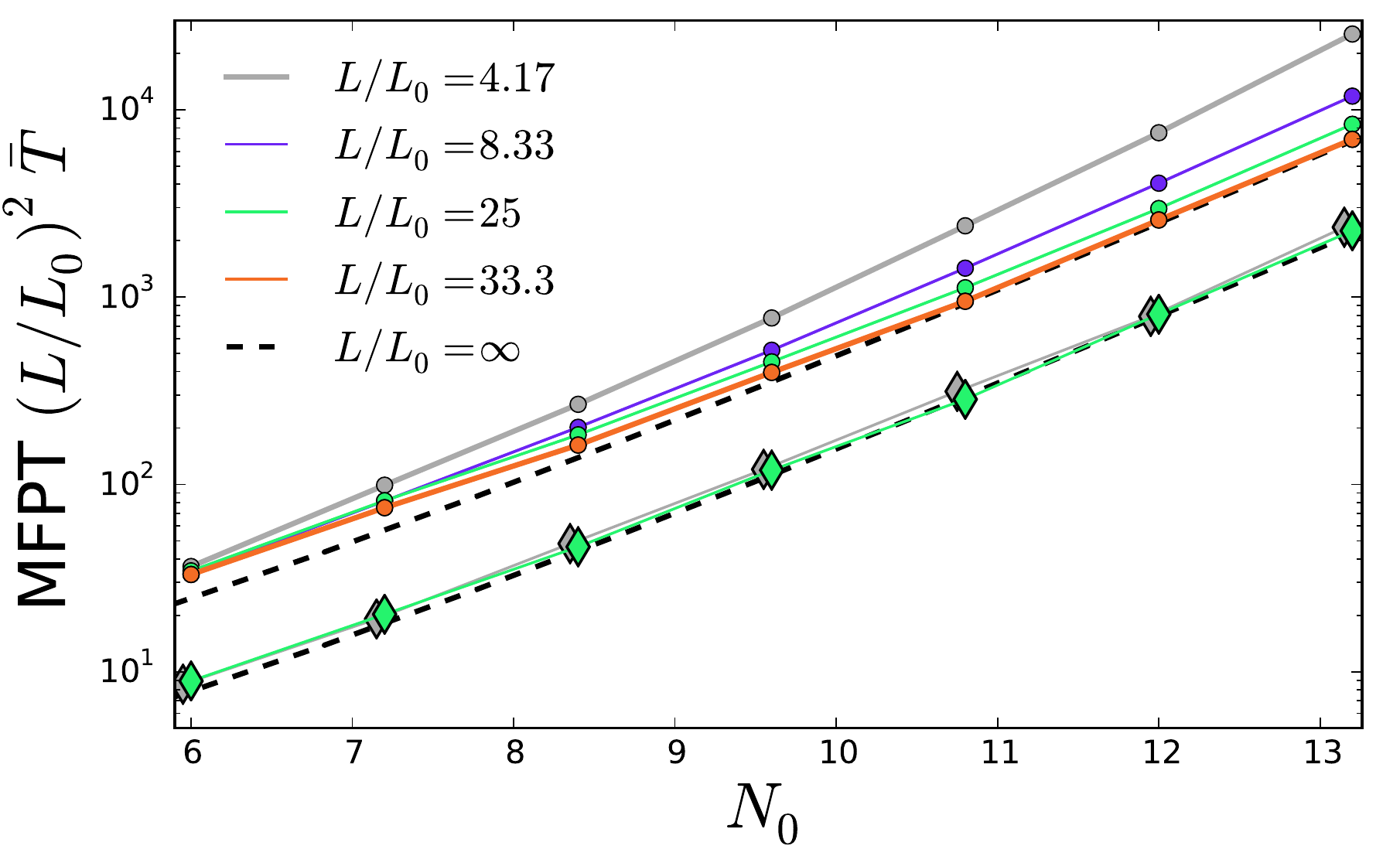}
  \caption{The MFPT vs $N_{0}$, the average number of particles in the ROI.  The symbols indicate $10^{3}$ averaged simulations; the 1D simulations are shown as circles and the 2D simulations are shown as diamonds. Also shown is the $L\to\infty$ approximation (dashed curve) for both 1D and 2D. Note that time is nondimensional using the $\dl^{2}/D$ timescale.}
  \label{fig:density}
\end{figure}

Finally, we  consider the case the radius of the ROI $\dl$ is fixed and $L\to \infty$ with a fixed number of particles per unit length $\varphi = N/(2L)$.
Let $N_{0}$ be the average number of particles in the ROI.
The $L\to\infty$ MFPT approximation (in dimensional units) is
\begin{equation}
\label{eq:7}
  \dT_{\infty} \sim \frac{\kappa_{\rm 1D}\dl^{2} e^{N_{0}}}{N_{0}^{2} D}, \quad N_{0} \gg 1.
\end{equation}
The MFPT is shown in Fig.~\ref{fig:density} as functions of $N_{0}$, for different values of $L$.
The approximation \eqref{eq:7} matches closely with the $L/\dl=33.3$ simulations for $N_{0} >10$.
Hence, the effect of a small domain size compared to the ROI is to increase the MFPT, making the cavitation event more rare.
This provides a quantitative measure of when the domain size $L$ no longer influences the cavitation event, which is relevant when, for example, considering cavitation on a relatively long strand of DNA compared to a shorter plasmid.
Microtubule filaments also vary in length.

Simulations of 2D cavitation are also shown in Fig.~\ref{fig:density} as diamond symbols.
Although the asymptotic approximation \eqref{eq:7} is derived for 1D, we find that it is a remarkably good fit to the simulation data after changing a single parameter: the prefactor $\kappa_{\rm 2D} \approx 0.7$.
We therefore infer that cavitation is roughly three times faster in 2D than in 1D.
Heuristically, this speed-up occurs because the mean time for a random walker to escape a spherical region decreases with dimensionality.

The asymptotic approximations \eqref{eq:3}-\eqref{eq:7} for 1D cavitation are derived as follows.
Because all of the $N$ walkers are independent, the problem can be reformulated as the first passage time of a single random walker in a $N$ dimensional domain.
Define the domain $\Omega \equiv (0, 1)^{N}$, and let $\Omega_{\epsilon}\equiv (1-\epsilon, 1)^{N}$ be the small target domain. 
The random process $\mathbf{R}(t) \in \Omega \setminus \Omega_{\epsilon}$ represents the original process with $\mathbf{R}(t) = (R_{1}(t),\cdots, R_{N}(t))$.
Define the MFPT as $T\equiv \ave{\inf \{t>0: \min_{1\leq n \leq N}R_{n}(t) = \ndl \}}$; it satisfies
\begin{gather}
  \label{eq:20}
    \sum_{n = 1}^{N}\pdd{T}{r_{n}} = -1, \quad \vr \in \Omega, \\
  \partial_{\bm{\eta}}T(\vr) = 0, \quad \vr \in \partial \Omega, \\
  T(\vr) = 0, \quad \vr \in \partial \Omega_{\epsilon}.
\end{gather}
An approximate solution to \eqref{eq:20} can be obtained using the method of matched asymptotics \cite{ward93a,condamin07c,schuss07a,cheviakov11a,isaacson13a}.
We split the solution into two parts: an inner and outer solution. 
The inner solution satisfies the absorbing boundary condition on $\partial\Omega_{\epsilon}$ and ignores the reflecting boundary.
The outer solution satisfies the reflecting boundary on $\partial \Omega$ and is singular as $\vr \to (1, \cdots, 1)$.
The two solutions are then matched to obtain a uniformly accurate approximation using the Van--Dyke matching principle \cite{keener00c}.

Define the inner coordinates $\vz = \frac{\vr - \vr_{b}}{\epsilon}$, and let $z = \norm{\vz}$.
The inner solution satisfies
\begin{equation}
  \label{eq:42}
  \Delta_{\vz} w = 0, \quad w(\vz\in \partial \mathcal{Z}_{N}) = 0,
\end{equation}
where $\mathcal{Z}_{N}$ is the unit hypercube.
The exact solution to the inner problem for arbitrary $N$ is unknown.
However, from electrostatics \cite{jackson62a}, for large $z$, the inner solution has the two term expansion,
\begin{equation}
  \label{eq:19}
  w \sim B_{N}(\epsilon)\left[\left(z/\CN\right)^{2-N} - 1\right], \quad N\geq 3.
\end{equation}
where $B_{N}$ is a constant determined by matching to the outer solution. 
The constant $\CN$, called the Newtonian capacitance, is a boundary dependent term discussed below.

Up to an unknown constant $\tilde{T}$, the outer solution is 
\begin{equation}
  \label{eq:36}
  T_{\rm out} \sim -G_{N}(\vr, \vr_{b}) + \tilde{T},
\end{equation}
where the Green's function $G_{N}$ satisfies,
\begin{gather}
  \label{eq:21}
    \sum_{n=1}^{N}\pdd{G_{N}}{r_{n}} = 1 - \delta(\vr - \vr'), \quad \vr \in \Omega, \\
  \partial_{\bm{\eta}}G_{N}(\vr, \vr') = 0, \quad \vr \in \partial \Omega, \\
\label{eq:6}
  \int_{\Omega}G_{N}(\vr, \vr')d\vr = 0.
\end{gather}
By integrating \eqref{eq:36} over $\Omega$ using \eqref{eq:6}, we find that $\tilde{T}$ is the MFPT averaged over a uniformly distributed set of initial positions, i.e., $\tilde{T}=\ndT$.
Again from electrostatics \cite{jackson62a}, in the limit $r_{n} \to 1$ with $r'_{n}=1$ and $\norm{\vr - \vr'} = \epsilon z$, the Green's function scales like
\begin{equation}
  \label{eq:5}
  G_{N} \sim 2^{N}\AN(\epsilon z)^{2-N} + O(1), \quad N \geq 3,
\end{equation}
where $\AN$ is given by \eqref{eq:2}.

Matching the inner and outer solutions we find that the $z$ dependent terms match provided that $B_{N}(\epsilon)=-2^{N}\AN\epsilon^{2-N}$ for $N\geq 3$.
The remaining unknown term $\ndT$ yields the approximation \eqref{eq:3}.

In order to access the rare event limit where both $N\gg 1$ and $\epsilon \ll 1$, we must find how the Newtonian capacitance $\CN$ scales with $N$.
This problem has no known exact solution for $N>2$ \cite{hwang10a}.

If the cuboid boundary $\partial \Omega_{\epsilon}$ were replaced by a spheroid with the same hypervolume, then the Newtonian capacitance is known for general $N$, 
\begin{equation}
\label{eq:80}
  \CN \approx \frac{2}{\sqrt{\pi}}\Gamma\left(1+\frac{N}{2}\right)^{1/N} \sim \sqrt{\frac{2N}{\pi e}}.
\end{equation}
We therefore propose a general expansion of $\CN$ (for the present case of cuboid boundary) having the same form as the large-$N$ expansion of \eqref{eq:80}, 
\begin{equation}\label{eq:75jun}
  \CN \sim \sqrt{\frac{2N}{\pi e}}\left(1 + \frac{\alpha_{1} \log N}{N} + \frac{\alpha_{2} }{N} + O(N^{-2})\right).
\end{equation}
Note that \eqref{eq:80} and \eqref{eq:75jun} have the same leading-order term. 

To elucidate how the unknown constants $\alpha_{1,2}$ affect the large $N$ MFPT approximation, we use Stirling's formula, leading to
\begin{equation}
  \label{eq:4}
  \frac{2^{N}\AN}{\CN^{N-2}} \sim \frac{\kappa_{\rm 1D}}{N^{\beta}},\quad N\gg 1,
\end{equation}
where 
\begin{equation}
\label{eq:4bjun}
\beta =  \alpha_{1}  + 1/2, \quad \kappa_{\rm 1D} = \frac{2}{\sqrt{\pi}e^{\alpha_{2}+1}}.
\end{equation}
In dimensional units, the MFPT approximation is
\begin{equation}
  \label{eq:40}
  \dT \sim \frac{L^{2}\kappa_{\rm 1D} }{N^{\beta}D}\left(1 - \frac{\dl}{L}\right)^{2 - N}.
\end{equation}
We determine the value of $\alpha_1$ by exploiting a physical constraint as follows. 
As $L\rightarrow\infty$ with the density of particles $\varphi=N/(2L)$ held constant, the MFPT must converge to a finite value.
Substituting $L = N/(2\varphi)$ and $N_{0} = 2\dl \varphi$ into \eqref{eq:40} yields
\begin{equation}
  \label{eq:1}
  \dT \sim \frac{N^{2-\beta}\kappa_{\rm 1D} }{4 \varphi^{2} D}\left(1 - \frac{N_{0}}{N}\right)^{2 - N}.
\end{equation}
Since $\lim_{N\to\infty}\left(1 - \frac{N_{0}}{N}\right)^{2 - N} = e^{N_{0}}$, we must have that $\beta=2$ (and therefore $\alpha_1=3/2$) in order for \eqref{eq:1} to converge to a finite, nonzero value in the limit $(L,N)\rightarrow\infty$. 
We also find that $\beta = 2$ is supported by numerical simulations (see Supplementary Material).
The limiting result is the approximation Eq.~\eqref{eq:7}.

While the approximation matches well with simulations in 2D, a more systematic asymptotic analysis for the 2D case should be feasible.
For small $\epsilon$ and finite $L$, the leading order in \eqref{eq:3} holds in 2D. 
A notable feature of our 1D case is that there are no terms in the expansion between the leading order term and the $O(1)$ term, making our 1D approximation converge particularly fast. 
This feature is lacking in 2D, where there are other terms singular in $\epsilon$, therefore we expect this approximation to converge more slowly. 
In 2D for large $L$ at constant density, a different scaling between $N$ and $L$ prevents the approximation in \eqref{eq:1} from converging, necessitating an alternative strategy that will be the subject of future research.

Returning to the specific question of cell-cell contact at T cell interfaces, large diffusing molecules such as CD45 disfavor proximity between receptors and ligands on apposing cells.
These molecules have diffusion coefficients of $D\approx 0.1 \mum^2/s$ \cite{Rajani:2011jpa} and  density such that on average there are $N_{0}=30$ molecules in the 100-nanometer ROI \cite{Allard:2012gy}. 
The approximation \eqref{eq:7}, using the prefactor $\kappa_{\rm 2D}=0.7$ from the numerical fit to simulations, yields an estimate of $\mathcal{T} \approx 10^9$ seconds.
In contrast, the MFPT for a single particle to escape a circular domain is $\mathcal{T} = L_{0}^{2}/(4D) = 0.025$ seconds.
Since T cell receptor triggering occurs within seconds \cite{Dushek:2009hd}, the above calculation predicts that receptor-ligand binding must involve a mechanism faster than passive diffusion. We therefore suggest the alternative hypothesis that an active force drives receptor-ligand proximity \cite{Allard:2012gy}.
To obtain an empty ROI spontaneously in less than five seconds, we would require $N_{0} \leq 7$, corresponding to a four-fold dilution, which could be experimentally accessible. The biological system is complicated by interactions of large molecules within and between molecular species, lipid heterogeneity, and transient immobilization, all of which could be exploited to dynamically tune the rate of ligand binding and will be studied by expanding the present framework. 
\section{Acknowledgments}
JN was supported by a NSF-funded postdoctoral fellowship (NSF DMS-1100281, DMS-1462992). JA was supported by a NSF CAREER award (DMS-1454739). 

\appendix

\section{Simulation algorithm}

The exact simulation algorithm makes use of two solutions to the 1D diffusion equation.
Let $p_{a, r}(x, x_{0}, t)$ be solutions to
\begin{gather}
  \label{eq:5}
  \pd{p}{t} = \pdd{p}{x}, \quad 0< x,x_{0} < x_{a,r} \\
  \pd{p}{x} = 0, \quad x = 0 \\
  p(x, x_{0}, 0) = \delta(x-x_{0}),
\end{gather}
with two different right boundary conditions.
Let $p_{a}$ be the solution with an absorbing BC at $x_{a} = \ndl = 1-\epsilon$.
This solution is used to derive $f$,  the jump time distribution.
Let $p_{r}$ be the solution with a reflecting BC at $x_{r} = 1$.
This solution is used to generate the random position of each particle given a jump time.
In both cases, the solution is represented as an infinite series.
Two different series representations are derived for each solution: one that converges quickly for short times and one for long times.

Let 
\begin{equation}
  \label{eq:4}
  a_{n} = \frac{\pi}{\ndl}(n-1/2), \quad b_{n} = n\pi.
\end{equation}
For large times, we have
\begin{equation}
  \label{eq:78}
  p_{a}(x, x_{0}, t) = \frac{2}{\ndl}\sum_{n = 1}^{\infty}\cos(a_{n}x)\cos(a_{n}x_{0})e^{-a_{n}^{2}t},
\end{equation}
and
\begin{equation}
  \label{eq:79}
  p_{r}(x \mid x_{0}, t) = 1 + 2\sum_{n = 1}^{\infty}\cos(b_{n}x)\cos(b_{n}x_{0})e^{-b_{n}^{2}t}.
\end{equation}
For short times we have
\begin{multline}
  \label{eq:82}
  p_{a}(x, x_{0}, t) = \frac{1}{\sqrt{4\pi t}}\sum_{n=0}^{\infty}(-1)^{n}\left(e^{-\frac{(2\ndl n + (x + x_{0}))^{2}}{4t}} - e^{-\frac{(2\ndl (n+1) - (x + x_{0}))^{2}}{4t}}\right. \\
\left. + e^{-\frac{(2\ndl n + \abs{x-x_{0}})^{2}}{4t}} - e^{-\frac{(2\ndl (n+1) - \abs{x - x_{0}})^{2}}{4t}}\right)
\end{multline}
and
\begin{multline}
  \label{eq:81}
  p_{r}(x, x_{0}, t) = \frac{1}{\sqrt{4\pi t}}\sum_{n=0}^{\infty}\left(e^{-\frac{(2n+(x + x_{0}))^{2}}{4t}} + e^{-\frac{(2(n+1) - (x + x_{0}))^{2}}{4t}} \right.\\
\left.+ e^{-\frac{(2n + \abs{x-x_{0}})^{2}}{4t}} + e^{-\frac{(2(n+1) - \abs{x - x_{0}})^{2}}{4t}}\right)
\end{multline}

For short times, the first passage time density is
\begin{equation}
  \label{eq:87}
\begin{split}
  f(t \mid x_{0}) &=-\pd{}{x}p_{a}(\ndl, x_{0}, t) \\
 &= \frac{4\pi}{(4\pi t)^{3/2}}\sum_{n=0}^{\infty}(-1)^{n}\left\{
(\ndl(2n+1) + x_{0})e^{-\frac{(\ndl(2n+1) + x_{0})^{2}}{4t}} \right.\\
& \qquad\qquad\qquad\qquad\qquad \left. + (\ndl(2n+1) - x_{0})e^{-\frac{(\ndl(2n+1) - x_{0})^{2}}{4t}} \right \}
\end{split}
\end{equation}
with the cumulative distribution,
\begin{equation}
  \label{eq:83}
\begin{split}
  F(t \mid x_{0}) = 1 + \sum_{n=0}^{\infty}(-1)^{n}\left\{
{\rm erf}(\frac{\ndl(2n+1) + x_{0}}{\sqrt{4t}}) + {\rm erf}(\frac{\ndl(2n+1) - x_{0}}{\sqrt{4t}})
 \right \}
\end{split}
\end{equation}
For long times, the first passage time density and cumulative distribution are
\begin{gather}
  \label{eq:96}
  f(t \mid x_{0}) = -\frac{2}{\ndl}\sum_{n=1}^{\infty}(-1)^{n}a_{n}\cos(a_{n}x_{0})e^{-a_{n}^{2}t}, \\
   F(t \mid x_{0}) =1 +\frac{2}{\ndl}\sum_{n = 1}^{\infty}(-1)^{n}\cos(a_{n}x_{0})\frac{e^{-a_{n}^{2}t}}{a_{n}}.
\end{gather}

The jump time is sampled using a standard root finding algorithm.
Given a uniform random variable $U$, the jump time is the unique solution to
\begin{equation}
  \label{eq:17}
  F(t \mid x_{0}) - U = 0.
\end{equation}

The distribution $p_{r}$ can be sampled using a rejection method as follows.
A majoring function $C(x)$ must be chosen such that $C(x) > p(x \mid x_{0}, t)$ for all $x\in (0, 1)$.
A tentative value $X$ is sampled from the distribution
\begin{equation}
  \label{eq:8}
  P(x) = \frac{C(x)}{\int_{0}^{1} C(x) dx}.
\end{equation}
A second random variable is drawn according to $Y = C(X)U$, where $U$ is a unit uniform random variable.
If $Y > p_{r}(X \mid x_{0}, t)$, then the sample $X$ is rejected.
The procedure is repeated until a sample is accepted.

For the long time expansion \eqref{eq:79} we select $X$ to be a uniform random variable in $(0, 1)$ and set
\begin{equation}
  \label{eq:6}
  Y = \left(\frac{1 + e^{-\pi^{2}t}}{1- e^{-\pi^{2}t}}\right)U.
\end{equation}
For the short time expansion \eqref{eq:81} we select $X$ to be a normal random variable with mean $x_{0}$ and variance $\sqrt{2t}$.
Note that care must be taken to ensure that $0<X<1$.
In this case,
\begin{equation}
  \label{eq:7}
  Y = e^{-(X - x_{0})^{2}/(4t)}\frac{U}{\sqrt{\pi t}}.
\end{equation}

\subsection{2D simulations}
For 2D cavitation, the outer boundary is a square of side length $2L$.
This geometry allows us to reuse the jump propagator from the 1D algorithm to update positions.
The $x$ and $y$ coordinate of each particle are updated from separate samples of the 1D propagator $p_{r}$ as described in the previous section.
The jump times are generated from the 2D distribution of first passage times to the boundary of a circle.

The 2D first passage time problem is
\begin{gather}
  \label{eq:9}
  \pd{}{t}p(r, t \mid r_{0}) = \frac{1}{r}\pd{}{r}\left(r\pd{p}{r}\right),\\
  p(r, 0 \mid r_{0}) = \frac{\delta(r-r_{0})}{2\pi r_{0}}, \\
 p(\ndl, t \mid r_{0}) = 0.
\end{gather}
The solution can be written as an expansion in Bessel functions.
The solution is
\begin{equation}
  \label{eq:13}
    p(r, t \mid r_{0}) = \frac{2}{\ndl}\sum_{j=1}^{\infty}\frac{J_{0}(r\beta_{n})J_{0}(r_{0}\beta_{n})}{J_{1}(\ndl\beta_{n})^{2}}e^{-\beta_{n}^{2}t},
\end{equation}
where $\alpha_{n}$ are the roots of $J_{0}(\alpha_{n}) = 0$, and $\beta_{n} = \alpha_{n}/\ndl$.
The jump time density function is
\begin{equation}
  \label{eq:14}
  f(t \mid r_{0}) = -\pd{}{r}p(l, t \mid r_{0}) = 
\frac{2}{\ndl}\sum_{j=1}^{\infty}\frac{\beta_{n}J_{0}(r_{0}\beta_{n})}{J_{1}(\ndl\beta_{n})}e^{-\beta_{n}^{2}t},
\end{equation}
and the cumulative distribution is
\begin{equation}
  \label{eq:15}
  F(t \mid r_{0}) = 1 - \frac{2}{\ndl}\sum_{j=1}^{\infty}\frac{J_{0}(r_{0}\beta_{n})}{\beta_{n}J_{1}(\ndl\beta_{n})}e^{-\beta_{n}^{2}t}.
\end{equation}
We use a root finding method to sample the jump time.

\section{Parameter estimation}
\begin{figure}[htbp]
  \centering
  \includegraphics[width=12cm]{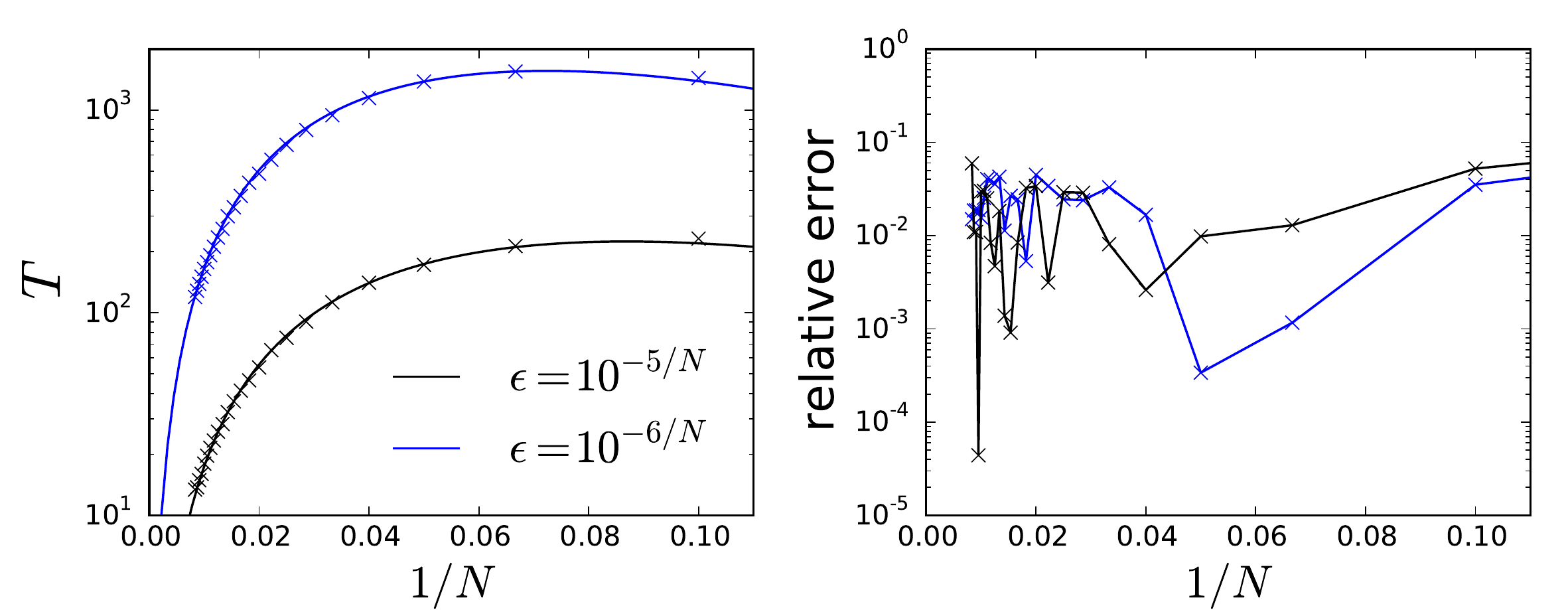}
  \caption{The MFPT approximation compared to Monte Carlo simulation estimates. Each symbol shows the sample mean of $10^{4}$ simulations.}
  \label{fig:1}
\end{figure}

\begin{figure}[htbp]
  \centering
  \includegraphics[width=12cm]{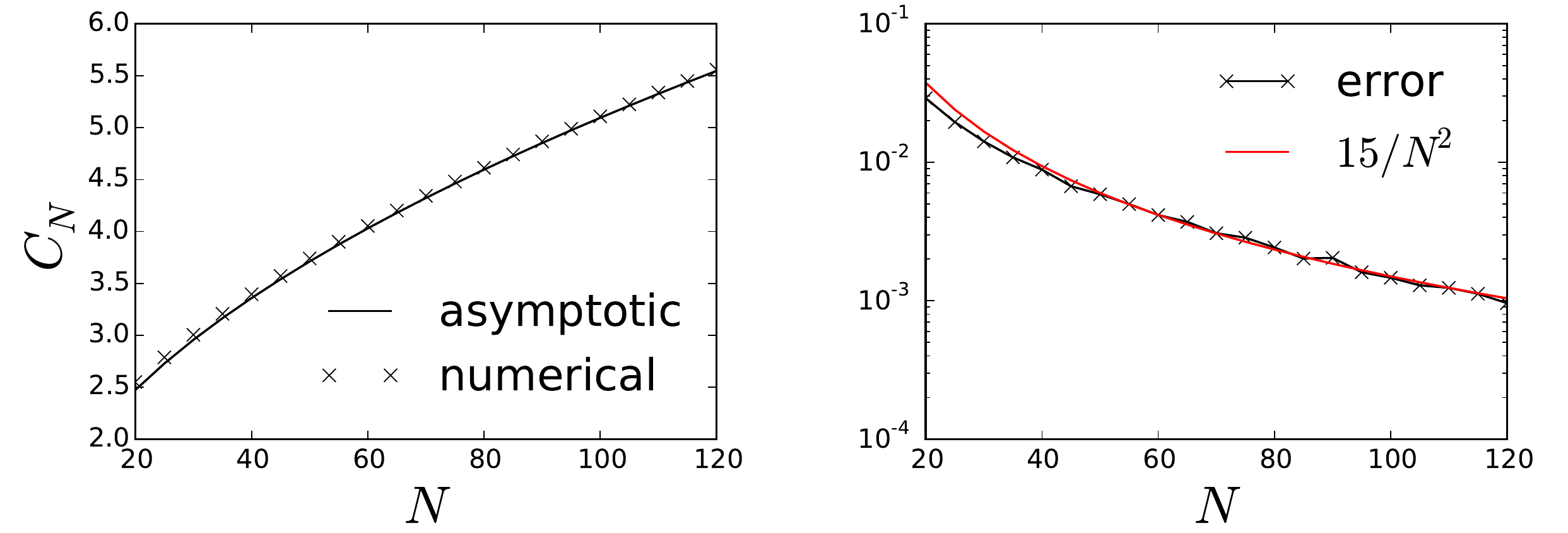}
  \caption{The Newtonian capacitance of a hypercube.}
  \label{fig:2}
\end{figure}

We use maximum likelihood to estimate parameter values in the large $N$ expansion of the Newtonian capacitance of a hypercube. 
We exploit the one to one correspondance between $\alpha_{1}$ and $\beta$ and between $\alpha_{2}$ and $\kappa$.
The parameters $\beta$ and $\kappa$ are estimated using realizations of the first passage time.
The likelihood function is computed by assuming that the first passage time is an exponentially distributed random variable with mean
\begin{equation}
  \label{eq:1}
  \ndT \sim \frac{\kappa }{N^{\beta}\epsilon^{N-2}}.
\end{equation}
This assumption is valid asymptotically as $N\to \infty$ when the first passage time is a rare event.
The likelihood function for $\beta$ and $\kappa$ from $n$ iid samples $\{\tau_{k}\}$, $k=1,\cdots, n$ is given by
\begin{equation}
  \label{eq:2}
  P( \{\tau_{k}\} \mid \beta, \kappa) = \exp\left[-n\left(\frac{T_{n}}{\ndT_{\beta,\kappa}} + \log \ndT_{\beta,\kappa}\right)\right],
\end{equation}
where
\begin{equation}
  \label{eq:3}
  T_{n} = \frac{1}{n}\sum_{k=1}^{n}\tau_{k}.
\end{equation}
Two data sets were generated for $N=20, 25, 30, \cdots, 115, 120$ with $\epsilon = 10^{-5/N}$ and $\epsilon = 10^{-6/N}$.
A value of $T_{n}$ was generated for each parameter set using $10^{4}$ samples of the first passage time from the exact simulation algorithm.
We numerically computed the maximum of the product of the likelihood functions from all parameter values.
The likelihood functions were computed on a $500 \times 500$ grid for $1.5 < \beta < 2.5$ and $1.5 < \kappa < 3$.
The resulting maximizers were $\beta \approx 2.00$ and $\kappa \approx 2.19$.
As shown in Fig.~\ref{fig:1}, $\ndT$ and $T_{n}$ are in good agreement with these parameter values.
The corresponding parameter values in the capacitance expansion are $\alpha_{1} \approx 3/2$ and $\alpha_{2} \approx -1.67$.
The capacitance approximation is shown in Fig.~\ref{fig:2} compared to numerical estimates.
From the expansion, we expect the error (given by the absolute difference divided by $\sqrt{N}$) to scale like $1/N^{2}$ as $N\to\infty$.
We find good agreement between the error and $15/N^{2}$.


\end{document}